\documentclass[rnaas,12pt]{aastex62}

\usepackage{natbib} 
\usepackage{graphicx}
\usepackage{enumitem}
\usepackage{lipsum}
\usepackage{ulem}
\usepackage{color}
\usepackage{url}

\usepackage{makecell}
\usepackage{longtable}

\usepackage{calc}
\newlength{\okinalen}
\newcounter{rownumber}

\begin{document}

\title{Evidence of Low-latitude Fluvial and Glacial Activity\\
During the Martian Amazonian Era}

\author[0000-0001-5702-2786]{Michael F. Zeilnhofer}
\affiliation{Department of Physics \& Astronomy, Northern Arizona University, PO Box 6010, Flagstaff, AZ 86011-6010, USA}

\author[0000-0001-7335-1715]{Colin Orion Chandler}
\affiliation{Department of Physics \& Astronomy, Northern Arizona University, PO Box 6010, Flagstaff, AZ 86011-6010, USA}

\author[0000-0002-5588-294X]{Nadine G. Barlow}
\affiliation{Department of Physics \& Astronomy, Northern Arizona University, PO Box 6010, Flagstaff, AZ 86011-6010, USA}









\section*{}
The Martian obliquity cycle is predominately influenced by Solar and Jovian tidal forces. The present-day axial tilt of Mars (25$^\circ$) is predicted to cycle between 0$^\circ$ and 60$^\circ$ \citep{Mischna:2005da} with excursions up to 80$^\circ$ \citep{Laskar:2004ie}. The obliquity cycle occurs over roughly 10$^{5}$--10$^{6}$ years \citep{Laskar:2002gq}, with recent models showing a $\sim$35$^\circ$ obliquity during the last 250 Myr \citep{Laskar:2004ie}. During an obliquity cycle, Mars is subjected to extreme climate change caused by rapid shifts in latitude-dependent solar insolation \citep{Forget:2006ef}. \cite{Forget:2006ef} probed this effect using a Global Circulation Model (GCM) to simulate  evolution of the Martian atmosphere, including potential southward polar ice cap migration at obliquities between 35$^\circ$ - 45$^\circ$.

\begin{figure}[ht]
\centering
\includegraphics[width=1.0\linewidth]{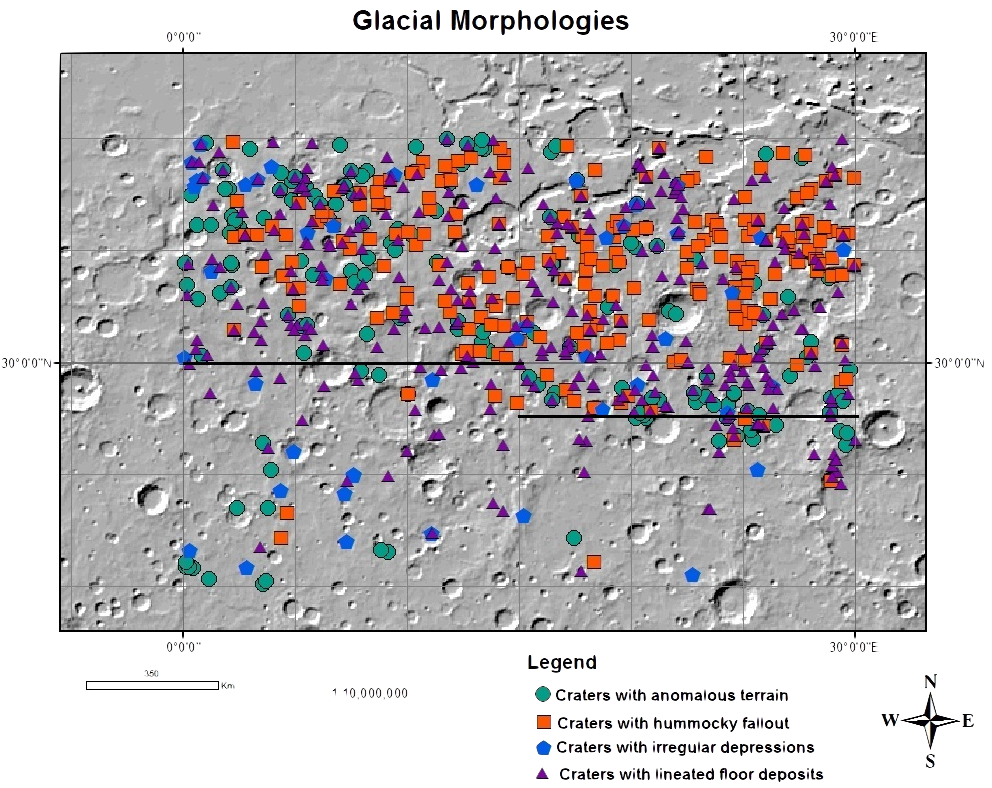}
\caption{This ArcGIS map details the distribution of all four glacial morphologies found within our region of study (20-40$^\circ$N 0-30$^\circ$E): anomalous terrain, hummocky fallout, irregular depressions, and lineated floor deposits. The black line indicates the boundary where these glacial features extend down to $\sim28^\circ$N.}
\label{fig:map}
\end{figure}


We focused on Arabia Terra, a region especially well-suited for studying glacial migration due to its location in the northern mid-latitude (20$^\circ$-40$^\circ$N, 0$^\circ$-30$^\circ$E) zone, and its heavily cratered terrain. These craters can serve as catch basins for glacial deposits. To analyze geologic features found within Arabia Terra we made use of Mars Reconnaissance Orbiter (MRO) and Mars Odyssey spacecraft imagery 
along with the \cite{Zeilnhofer:2015vi} small crater database and the \cite{Barlow:1988kq,Barlow:2017wg} large crater databases. Spacecraft data were collected with the MRO Context Camera (6 m/pixel; \citet{Malin:2007hn}) and the Mars Odyssey Thermal Emission Imaging System (THEMIS; 100 m/pixel daytime IR, 18 m/pixel visible; \citet{Christensen:2004bo}). We selected craters with interior morphologies of glacial origin, fluvial origin, or a combination of the two. Glacial morphologies in our dataset include anomalous terrain, hummocky fallout, irregular depressions, and lineated floor deposits. Fluvial morphologies include channel deposits, floor channels, and layered deposits.

We created glacial and fluvial morphology distribution maps with the ESRI\footnote{\url{https://www.esri.com/en-us/home}} Aeronautical Reconnaissance Coverage Geographic Information System (ArcGIS). Figure \ref{fig:map} shows the distribution of glacial morphologies in North-central Arabia Terra. We found glacial deposits across all longitudes of our study, and as far south as $\sim$28$^\circ$N. The fluvial deposits were evenly distributed throughout.

We determined the ages of our craters in order to establish a time-domain link between the Martian obliquity cycle and glacial migration. We employed the \textit{CraterStats} \citep{Michael:2010ke} software package to produce size-frequency distribution (SFD) plots. We used both chronologies available to \textit{CraterStats}: the G. Neukum \citep{Neukum:1994ub,Ivanov:2001gj} and W. Hartmann \citep{Hartmann:2001cq} chronologies. The average age for the four glacial deposits in our dataset was 2.27$\pm$0.42 Ga (via Hartmann) or 1.99$\pm$0.37 Ga (via Neukum). In computing the glacial deposit ages we selected highly degraded (ancient) craters to determine an upper limit and the freshest craters to establish a lower limit (when glacial activity subsided). We arrived at a lower bound of 43.3$\pm$9.8 Ma (Hartmann) or 56.6$\pm$13 Ma (Neukum). We noted fluvial morphologies were typically younger than glacial morphologies by both chronologies: 2.10 $\pm$0.51 Ga (Hartmann) and 1.80 $\pm0.45$ Ga (Neukum). Our computed ages are in agreement with both the GCM of \cite{Forget:2006ef} and the obliquity model of \citep{Laskar:2004ie}, and these younger ages suggest some of the fluvial activity could have been caused by glacial melting.

\textit{Acknowledgements:} The authors wish to thank Prof. Ty Robinson and Lauren Biddle of Northern Arizona University whose comments greatly improved the quality of this research note, and Dr. Lisa Prato (Lowell Observatory) for inspiring this submission. This research was supported by NASA Grant 1000688 to NGB.

\newpage
\bibliography{papers.bib}

\end{document}